\documentclass[letterpaper, 10 pt, conference]{ieeeconf}  

\IEEEoverridecommandlockouts                              
\usepackage{cite}
\usepackage{amsmath,amssymb,amsfonts,bm}
\usepackage{bbm}
\usepackage{graphicx}
\usepackage{subfigure}
\usepackage{textcomp}
\usepackage{algorithm}
\usepackage{algpseudocode}

\usepackage{xcolor}
\newcommand{\tabincell}[2]{\begin{tabular}{@{}#1@{}}#2\end{tabular}}
\usepackage{mathrsfs}
\def \qed {\hfill \vrule height6pt width 6pt depth 0pt}
\allowdisplaybreaks[4]

\newtheorem{lemma}{Lemma}
\newtheorem{assum}{Assumption}

\newtheorem{remark}{Remark}

\title{\LARGE \bf
	Regularization for Covariance Parameterization of   Direct Data-Driven LQR Control
}

\author{Feiran Zhao, Alessandro Chiuso, Florian D\"{o}rfler
		\thanks{F.~Zhao and F. D\"{o}rfler are with the Department of Information Technology and Electrical Engineering, ETH Z\"{u}rich, 8092 Z\"{u}rich, Switzerland. (e-mail: zhaofe@control.ee.ethz.ch; dorfler@control.ee.ethz.ch) A. Chiuso is with the Department of Information Engineering, University of Padova, Via Gradenigo 6/b, 35131 Padova, Italy. (e-mail: alessandro.chiuso@unipd.it)}}%
\begin{document}
	
	\maketitle
	\thispagestyle{empty}
	\pagestyle{empty}
	
	\begin{abstract}
	As the benchmark of data-driven control methods, the linear quadratic regulator (LQR) problem has gained significant attention. A growing trend is direct LQR design, which finds the optimal LQR gain directly from raw data and bypassing system identification. To achieve this, our previous work develops a direct LQR formulation parameterized by sample covariance. In this paper, we propose a regularization method for the covariance-parameterized LQR. We show that the regularizer accounts for the uncertainty in both the steady-state covariance matrix corresponding to closed-loop stability, and the LQR cost function corresponding to averaged control performance. With a positive or negative coefficient, the regularizer can be interpreted as promoting either exploitation or exploration, which are well-known trade-offs in reinforcement learning. In simulations, we observe that our covariance-parameterized LQR with regularization can significantly outperform the certainty-equivalence LQR in terms of both the optimality gap and the robust closed-loop stability. 
	\end{abstract}
	
	
	\section{Introduction}
	As a cornerstone of modern control theory, the linear quadratic regulator (LQR) has become \textit{the} benchmark problem of validating and comparing different data-driven control methods \cite{dean2020sample}. Manifold approaches to data-driven LQR design can be broadly classified as indirect, i.e., based on system identification (SysID) followed by model-based control, versus direct when by-passing SysID~\cite{dorfler2021certainty}. 
	 
	Classical indirect LQR design is based on certainty-equivalence principle: a system is first identified from data; then, the LQR is obtained with Riccati equations by treating the estimated system as the ground-truth\cite{mania_certainty_2019}. When identifying models from data, regularization methods can be used for the regression problem to deal with uncertainties and facilitate numerical computation~\cite{pillonetto2022regularized}. The indirect certainty-equivalence LQR design is widely studied and has been proved to achieve optimal regret in online adaptive control~\cite{simchowitz2020naive}. There are also indirect adaptive control methods \cite{cohen2019learning, abeille2020efficient, chekan2024fully} adding a covariance-based regularizer in the cost function to promote exploration.
	 
 	Motivated from behavioral system theory and subspace methods, the direct LQR design has become a growing trend  in data-driven control field \cite{markovsky2021behavioral}. The seminal works \cite{de2019formulas, van2020data} propose a data-based system parameterization, where the state-feedback gain is parameterized as a linear function of a batch of persistently exciting data. Leveraging subspace relations, the closed-loop matrix can be further expressed by raw data matrices. As such, the LQR problem can be reformulated as a data-based convex program parameterized and solved without involving any explicit SysID. In this direct LQR framework, regularization can be used to single out a solution with favorable properties~\cite{de2021low,dorfler2021certainty,dorfler22on}. By selecting proper regularization coefficients, the solution can flexibly interpolate between indirect certainty-equivalence LQR and robust closed-loop stable gains.
 	
 	While the parameterization and regularization in \cite{de2021low,dorfler22on,dorfler2021certainty} sheds a light on direct LQR design,  two limitations hinder their broader implication. First, the dimension of their direct LQR formulation scales linearly with the data length. Thus, this parameterization cannot be used to achieve adaptive control with online closed-loop data~\cite{zhao2024data}. Second,
 	their direct LQR solution without regularization is sensitive to noise \cite{zeng2024noise}. Even using regularization, there is a trade-off between \textit{performance} and \textit{robust closed-loop stability} in their solution, i.e., one has to be sacrificed to gain the other \cite{dorfler22on}. Moreover, when the length of data tends to infinity, the regularized formulation may lead to trivial solutions~\cite{zeng2024noise}.

 	To address the first limitation, our previous work \cite{zhao2024data} proposes a new parameterization for the direct data-driven LQR, which parameterizes the feedback gain as a linear function of the sample covariance of input-state data. A key distinction of this covariance parameterization is that the dimension of the direct LQR formulation does not scale with the data length, which enables direct adaptive control using data-enabled policy optimization (DeePO)~\cite{zhao2023data}. 
 	Without using any regularization, the covariance-parameterized LQR is shown to be equivalent to the indirect certainty-equivalence LQR, which has a certain degree of robustness against noise. Recently, the covariance parameterization and the DeePO method have been applied for power converter system~\cite{zhao2024direct} and autonomous bicycle~\cite{persson2025adaptive}.
 	
%
 	Recently, there have been regularization methods accounting for the uncertainty in data-driven predictive control~\cite{chiuso2023harnessing, grimaldi2024bayesian}. In particular, the work \cite{chiuso2023harnessing} shows that a separation principle holds, i.e., the expectation of the cost function given data can be decoupled by the sum of the certainty-equivalence cost and the uncertainty linear in the covariance of the model estimator. By introducing a covariance-based regularizer, the uncertainty in the cost is well compensated, leading to significant improvement in control performance.

 	In this paper, we propose a regularization method for the covariance-parametrized LQR~\cite{zhao2024data}. We show that the regularizer accounts for the uncertainty in both the steady-state covariance matrix corresponding to closed-loop stability, and the LQR cost function corresponding to averaged control performance. Hence, it resolves the trade-off between robust closed-loop stability and performance observed in \cite{dorfler2021certainty,dorfler22on}.  Consistent with \cite{chiuso2023harnessing, grimaldi2024bayesian}, our regularizer is also linear in the covariance matrix of the model estimator in terms of indirect LQR control. With a positive or negative coefficient, the regularizer can promote either exploitation or exploration, which are well-known trade-offs in reinforcement learning~\cite{recht2019tour}. In particular, the negative regularizer coincides with that in indirect adaptive control used for exploration \cite{cohen2019learning, abeille2020efficient, chekan2024fully}.

 	
 	In simulations, we validate the effectiveness of the regularization method on a benchmark problem \cite{dean2020sample}. As a remarkable empirical result,  
 	our covariance-parameterized LQR with properly tuned regularization can significantly outperform the  certainty-equivalence LQR in terms of both the optimality gap and the probability of obtaining a stabilizing controller. 
 	
 	The remainder of this paper is organized as follows. Section \ref{sec:prob} recapitulates data-driven formulations of the LQR problem. Section \ref{sec:regu} proposes the regularization method for direct data-driven LQR control. Section \ref{sec:simu} validates the results via simulations. Conclusions are made in Section \ref{sec:conc}.
 	
 	\textbf{Notation.} We use $I_n$ to denote the $n$-by-$n$ identity matrix. We use $\underline{\sigma}(\cdot)$ to denote the minimal singular value of a matrix.  We use $A^\dagger:=A^{\top}(AA^{\top})^{-1}$ to denote the right inverse of a full row rank matrix $A\in \mathbb{R}^{n\times m}$. We use $\mathcal{N}(A)$ to denote the nullspace of $A$. We use $o(f(t))$ to denote a function $g(t)$ that satisfies the condition $\lim\limits_{t\rightarrow \infty} (g(t)/f(t)) = 0$.

	\section{Data-driven formulations of the LQR}\label{sec:prob}
	In this section, we recall the model-based LQR, indirect certainty-equivalence LQR \cite{dorfler22on}, and direct LQR design using data-based parameterization \cite{de2019formulas,de2021low, zhao2024data}.
	\subsection{Model-based LQR}
	Consider a discrete-time linear time-invariant (LTI) system
	\begin{equation}\label{equ:sys}
		x_{t+1} = Ax_t + Bu_t + w_t,
	\end{equation}
	where $t\in \mathbb{N}$, $x_t\in\mathbb{R}^{n}$ is the state,   $u_t\in\mathbb{R}^{m}$ is the control input, $\{w_t\}$ is random process noise, and $(A,B)$ is stabilizable. Throughout the paper, we make the following assumption on the noise.
	\begin{assum}
		\label{assumption:noise}
		The noise sequence $\{w_t\}$ is identically and independently distributed with $w_t \sim \mathcal{N}(0, I_n)$\footnote{While we assume a unit covariance of noise for the simplicity of presentation, all the results can be extended to general covariance case straightforwardly.}.
	\end{assum}

	The infinite-horizon LQR problem aims to find a state feedback gain $K$ that minimizes a time-average cost, i.e.,
	\begin{equation}\label{prob:lqr}
		\begin{aligned}
			&\text{minimize}_{K} ~~ \lim\limits_{T \rightarrow  \infty} \frac{1}{T} \mathbb{E}\left[\sum_{t=0}^{T-1}(x_{t}^{\top} Q x_{t}+u_{t}^{\top} R u_{t})\right]\\
			&\text{subject to} ~~~~~~~(\ref{equ:sys})~\text{and}~u_t = Kx_t,
		\end{aligned}
	\end{equation}
	where $Q\succ 0, R\succ 0$ are penalty matrices, and the expectation is taken over the statistics of noise.

 
	When $A+BK$ is stable, it holds that \cite{anderson2007optimal}
	\begin{equation}\label{equ:transfer}
		C(K) = \text{Tr}((Q+K^{\top}RK)\Sigma),
	\end{equation}
	where $\Sigma$ is the  steady-state covariance matrix obtained as the positive definite solution to the Lyapunov equation
	\begin{equation}\label{equ:Sigma}
		\Sigma = I_n + (A+BK)\Sigma (A+BK)^{\top}.
	\end{equation}
	We refer to $C(K)$ as the LQR cost and to (\ref{equ:transfer})-(\ref{equ:Sigma}) as a \textit{parameterization} of the LQR.
	
	The optimal LQR gain $K^*:=\arg\min_{K}C(K)$ of (\ref{equ:transfer})-(\ref{equ:Sigma}) can be found by the celebrated Riccati equation with known $(A,B)$ \cite{anderson2007optimal}. In the sequel, we recapitulate several data-driven control methods to find $K^*$ with unknown $(A,B)$.


	
	%

	\subsection{Indirect certainty-equivalence LQR with ordinary least-square identification}
	A classical approach to data-driven LQR design is based on the certainty-equivalence principle: it first estimates a nominal model $(A,B)$ from data, and then solves the LQR problem regarding the identified model as the ground-truth. Consider a $t$-long time series\footnote{The time series do not have to be consecutive. All results in Sections \ref{sec:prob}-\ref{sec:regu} also hold when each column in (\ref{equ:data}) is obtained from independent experiments or even averaged data sets.} of states, inputs, noises, and successor states
	\begin{equation}\label{equ:data}
	\begin{aligned}
	X_{0} &:= \begin{bmatrix}
	x_0& x_1& \dots& x_{t-1}
	\end{bmatrix}\in \mathbb{R}^{n\times t},\\
	U_{0} &:= \begin{bmatrix}
	u_0& u_1& \dots& u_{t-1}
	\end{bmatrix}\in \mathbb{R}^{m\times t}, \\
	W_{0} &:= \begin{bmatrix}
	w_0& w_1& \dots& w_{t-1}
	\end{bmatrix}\in \mathbb{R}^{n\times t}, \\
	X_{1} &:= \begin{bmatrix}
	x_1& x_2& \dots& x_t
	\end{bmatrix}\in \mathbb{R}^{n\times t},
	\end{aligned}
	\end{equation}
	which satisfy the system dynamics
	\begin{equation}\label{equ:dynamics}
	X_1 = AX_0+ BU_0 + W_0.
	\end{equation}
	
	We assume that the data is {\em persistently exciting (PE)} \cite{willems2005note}, i.e., the block matrix of input and state data 
	$D_0 := 	[U_0^{\top}, X_0^{\top}]^{\top}$  
	has full row rank
	\begin{equation}\label{equ:rank}
	\text{rank}(D_0) = m+n.
	\end{equation}
	It is well-known that this PE condition is necessary for the data-driven LQR design~\cite{van2020data,kang2023minimum}.
	
	Based on the subspace relations \eqref{equ:dynamics} and the rank condition (\ref{equ:rank}), the least-squares estimator $(\widehat{A},\widehat{B})$ of the system is  
	\begin{equation}\label{equ:sysid}
	[\widehat{B}, \widehat{A}] = \underset{B, A}{\arg \min }\left\|X_1-[B,A] D_0\right\|_F = X_1D_0^{\dagger}.
	\end{equation}
	It is an unbiased estimator with variance \cite[Chapter 3]{pillonetto2022regularized}
	\begin{equation}\label{equ:variance}
	\mathbb{E}_{W_0}\left[\begin{bmatrix}
		\widehat{B}-B &  \widehat{A}-A
	\end{bmatrix}\begin{bmatrix}
		\widehat{B}-B &  \widehat{A}-A
	\end{bmatrix}^{\top}\right] = \frac{1}{t}\Phi^{-1},
	\end{equation}
	where $\Phi :  = D_0D_0^{\top}/t$ is the sample covariance of input-state data.
	Following the certainty-equivalence principle~\cite{dorfler2021certainty}, the system $(A,B)$ is replaced with its estimate $(\widehat{A},\widehat{B})$ in (\ref{equ:transfer})-(\ref{equ:Sigma}), and the LQR problem can be reformulated as 
	\begin{equation}\label{prob:indirect}
	\begin{aligned}
	\mathop{\text{minimize}}\limits_{K, \Sigma\succeq 0} ~~&\text{Tr}\left((Q+K^{\top}RK)\Sigma\right),\\
	\text{subject to} ~~&\Sigma = I_n + (\widehat{A} + \widehat{B}K)\Sigma (\widehat{A} + \widehat{B}K)^{\top}\\
	&[\widehat{B}, \widehat{A}] = \underset{B, A}{\arg \min }\left\|X_1-[B,A] D_0\right\|_F.
	\end{aligned}
	\end{equation}
	We refer to  \eqref{prob:indirect} as the indirect certainty-equivalence LQR problem. 
	
	

	\subsection{Direct LQR with data-based system parameterization}
	 The{ \em direct data-driven} LQR design aims to find $K^*$ bypassing system identification  \eqref{equ:sysid}~\cite{de2019formulas,de2021low,dorfler2021certainty}. The key is the data-based parameterization: by the rank condition \eqref{equ:rank}, there exists $G \in \mathbb{R}^{t\times n}$ such that
	\begin{equation}\label{equ:relation}
	\begin{bmatrix}
	K \\
	I_n
	\end{bmatrix}=
	D_0G
	\end{equation}
	for any given $K$. Together with the subspace relation (\ref{equ:dynamics}), the closed-loop system matrix can be written as
	$$
	A+BK=[B,A]\begin{bmatrix}
	K \\
	I_n
	\end{bmatrix}\overset{\eqref{equ:relation}}{=}[B,A]D_0G\overset{\eqref{equ:dynamics}}{=}(X_1 - W_0)G,
	$$
	Following the certainty-equivalence principle~\cite{dorfler2021certainty}, the unknown noise matrix $W_0$ is disregarded, and $X_1G$ is used as the closed-loop matrix. Substituting $A+BK$ with $X_1G$ in (\ref{equ:transfer})-(\ref{equ:Sigma}) and together with \eqref{equ:relation}, the LQR problem becomes 
	\begin{equation}\label{prob:equi}
	\begin{aligned}
	&\mathop{\text {minimize}}\limits_{G, \Sigma\succeq 0}~ \text{Tr}\left((Q+G^{\top}U_0^{\top}RU_0G)\Sigma\right),\\
	&\text{subject to}~~ \Sigma = I_n + X_1G\Sigma G^{\top}X_1^{\top},~X_0G =I_n 
	\end{aligned}
	\end{equation}
	with the gain matrix $K = U_0G$, which can be reformulated as a semi-definite program (SDP) after changing variables~\cite{de2019formulas}. The LQR parameterization \eqref{prob:equi} is direct data-driven, as it does not involve any explicit SysID. Note that the dimension of \eqref{prob:equi} scales linearly with data length. 
	
	When noise is present, regularization methods are used to single out a solution with favorable properties. Three regularizers are developed in \cite{de2021low,dorfler2021certainty,dorfler22on}, including the certainty-equivalence promoting regularizer $\|\Pi_{D_0}G \|$, the robustness promoting regularizer $\text{Tr}(G\Sigma G^{\top})$, and the low-rank regularizer $\|G\|_1$. For the certainty-equivalence promoting regularizer with a sufficiently large coefficient, the solution coincides with the indirect certainty-equivalence LQR \eqref{prob:indirect}. A detailed discussion can be found in \cite{dorfler22on}.

	\subsection{Covariance parameterization of direct data-driven LQR}
	 Our previous work \cite{zhao2024data} proposes a sample covariance parameterization for the LQR problem. 
	Under the PE rank condition (\ref{equ:rank}), the sample covariance $\Phi$ is positive definite, and there is a \textit{unique} solution $V\in \mathbb{R}^{(n+m)\times n}$ to
	\begin{equation}\label{equ:newpara}
	\begin{bmatrix}
	K \\
	I_n
	\end{bmatrix}= \Phi V
	\end{equation}
	for any given $K$. We refer to \eqref{equ:newpara} as the \textit{covariance parameterization} of the policy. In contrast to the prior parameterization in (\ref{equ:relation}), the dimension of $V$ is independent of the data length. 
	
	\begin{remark}
	The solution of (\ref{equ:relation}) can be parameterized via orthogonal decomposition
	\begin{equation}\label{equ:rela}
		G = \frac{1}{t} D_0^{\top}V + \Delta, \Delta \in \mathcal{N}(D_0). 
	\end{equation}
    If we remove the nullspace $\mathcal{N}(D_0)$ in (\ref{equ:rela}), then the parameterization (\ref{equ:relation}) reduces to (\ref{equ:newpara}). This can be achieved by using the certainty-equivalence promoting regularizer $\|\Pi_{D_0}G \|$ for \eqref{prob:equi}~\cite{dorfler2021certainty}. \qed
	\end{remark}

	

	For brevity, define the sample covariance matrices
	 $\overline{X}_0= X_0D_0^{\top}/t$, $\overline{U}_0= U_0D_0^{\top}/t$, $\overline{W}_0= W_0D_0^{\top}/t$, and $\overline{X}_1= X_1D_0^{\top}/t.
	$
	Then, the closed-loop matrix can be written as
	\begin{equation}\label{equ:clos}
	A+BK=[B,A]\begin{bmatrix}
	K \\
	I_n
	\end{bmatrix}\overset{\eqref{equ:newpara}}{=}[B,A]\Phi V\overset{\eqref{equ:dynamics}}{=}(\overline{X}_1 - \overline{W}_0)V.
	\end{equation}
	Analogous to \eqref{prob:equi}, we disregard the uncertainty  $\overline{W}_0$ in the parameterized closed-loop matrix and formulate the direct data-driven LQR problem using $(X_0,U_0,X_1)$ as
	\begin{equation}\label{prob:equiV}
	\begin{aligned}
	&\mathop{\text {minimize}}\limits_{V, \Sigma\succeq 0}~J(V) :=\text{Tr}\left((Q+V^{\top}\overline{U}_0^{\top}R\overline{U}_0V)\Sigma\right),\\
	&\text{subject to}~ ~\Sigma = I_n + \overline{X}_1V\Sigma V^{\top}\overline{X}_1^{\top},\overline{X}_0V= I_n
	\end{aligned}
	\end{equation}
	with the gain matrix $K = \overline{U}_0V$ recovered from \eqref{equ:newpara}. We refer to (\ref{prob:equiV}) as the LQR problem with covariance parameterization. 
	It can be shown that  \eqref{prob:equiV} is equivalent to the indirect certainty-equivalence LQR (\ref{prob:indirect}) in the sense that their solutions coincide; see \cite[Lemma 1]{zhao2024data}. Thus, the certainty-equivalence promoting regularizer $\|\Pi_{D_0}G \|$ is not needed for the covariance parameterization \eqref{prob:equiV}.

	We show that the uncertainty $\overline{W_0}V$ in  \eqref{equ:clos} decreases with the amount of data. For brevity, define $\phi_t := [u_t^{\top}, x_t^{\top}]^{\top}$ and $\Phi_t := \mathbb{E}[\sum_{i=0}^{t-1}\phi_i\phi_i^{\top}/t]$.
	\begin{lemma}\label{lem:dec}
		Under Assumption \ref{assumption:noise}, it holds that $\mathbb{E}\left[\overline{W}_0\right] = 0$ and $\text{Var}\left[\text{vec}(\overline{W}_0)\right] = I_n \otimes \Phi_t/t$.
	\end{lemma}
	\begin{proof}
		 Since $w_t$ is uncorrelated with $\phi_t$, it holds that $\mathbb{E}\left[\overline{W}_0\right] = \mathbb{E}[\sum_{i=0}^{t-1}w_i\phi_i^{\top}/t]=0$. For the variance, we have
		 \begin{align*}
		 	&\text{Var}\left[\text{vec}(\overline{W}_0)\right]  = \text{Var}\left[  \text{vec}\left(\frac{1}{t}\sum_{i=0}^{t-1}w_i\phi_i^{\top}\right)\right]\\
		 	& = \frac{1}{t^2}\sum_{i=0}^{t-1}\sum_{k=0}^{t-1}\mathbb{E}[\text{vec}(w_i\phi_i^{\top})\cdot \text{vec}(w_k\phi_k^{\top})^{\top}] \\
		 	& = \frac{1}{t^2}\sum_{i=0}^{t-1}\sum_{k=0}^{t-1}\mathbb{E}[(w_i \otimes \phi_i)\cdot (w_k^{\top} \otimes \phi_k^{\top})] \\
		 	& = \frac{1}{t^2}\sum_{i=0}^{t-1}\sum_{k=0}^{t-1}\mathbb{E}[(w_i  w_k^{\top})\otimes (\phi_i  \phi_k^{\top})] \\
		 	& = \frac{1}{t^2}\sum_{i=0}^{t-1} I_n \otimes \mathbb{E}[\phi_i  \phi_i^{\top}] 
		 	= \frac{1}{t}I_n \otimes \Phi_t,
		 \end{align*}
		 where the third and fourth equalities follow from properties of Kronecker product.
	\end{proof}

	By Lemma \ref{lem:dec}, if $\phi_t$ converges to some steady distributions, i.e., $\Phi_t$ converges to some constant matrix, then the variance $\text{Var}\left[\text{vec}(\overline{W}_0)\right]$ and hence the uncertainty $\overline{W_0}V$ will decrease to zero with the rate $\mathcal{O}(1/t)$.  

%
	
	\section{Regularization for the covariance-parameterized LQR}\label{sec:regu}
	
	In this section, we first propose a regularization method for the covariance-parameterized certainty-equivalent LQR problem \eqref{prob:equiV} to enhance robust closed-loop stability\footnote{We use robust closed-loop stability to describe the robustness of the data-driven LQR  solution against noisy data, which can be quantified by the percentage of stabilizing solutions from multiple trails.}. Then, we show that this regularizer can also account for the uncertainty in the cost function. Finally, we interpret the regularizer as promoting exploration or exploitation.
	
	\subsection{Regularization for promoting robust closed-loop stability}
	
	The feasibility of the covariance-parameterized LQR problem \eqref{prob:equiV} depends on that of the Lyapunov equation
	\begin{equation}\label{equ:lyap}
		\Sigma = I_n + \overline{X}_1V\Sigma V^{\top}\overline{X}_1^{\top},
	\end{equation}
	where $\overline{X}_1V$ is regarded as the closed-loop matrix. However, by the relation $A+BK = (\overline{X}_1 - \overline{W}_0)V$, the Lyapunov equation that should be met is
	\begin{equation}\label{equ:true_lyap}
		\Sigma = I_n + (\overline{X}_1 - \overline{W}_0)V\Sigma V^{\top}(\overline{X}_1 - \overline{W}_0)^{\top}.
	\end{equation}
	
	To render the resulted controller stabilizing, we need the difference between the right-hand side of \eqref{equ:lyap} and \eqref{equ:true_lyap} to be small, which is                                      
	\begin{equation}\label{def:diff}
		 \Sigma_{\text{diff}}:=\overline{W}_0 V\Sigma V^{\top} \overline{X}_1^{\top} +  \overline{X}_1V\Sigma V^{\top} \overline{W}_0^{\top} - 	\overline{W}_0 V\Sigma V^{\top}  \overline{W}_0^{\top}.
	\end{equation} 
	By the subspace relations \eqref{equ:dynamics}, the difference can be further expressed as
	\begin{equation}\label{equ:gap}
		\begin{aligned}
		\Sigma_{\text{diff}}&\overset{\eqref{equ:dynamics}}{=}\overline{W}_0 V\Sigma V^{\top} ([B~A]\Phi + \overline{W}_0)^{\top}\\
		& ~~~+  ([B~A]\Phi + \overline{W}_0)V\Sigma V^{\top} \overline{W}_0^{\top} - 	\overline{W}_0 V\Sigma V^{\top}  \overline{W}_0^{\top}\\
		&=\overline{W}_0V\Sigma V^{\top}\overline{W}_0+\overline{W}_0V\Sigma V^{\top}\Phi [B~ A]^{\top} \\
		&~~~+ [B~ A]\Phi V \Sigma V^{\top} \overline{W}_0.
		\end{aligned}
	\end{equation}
	By Lemma \ref{lem:dec}, it follows that $\|\overline{W}_0\| \sim 1/\sqrt{t}$ as $t$ grows to infinity. Hence, the first term  in \eqref{equ:gap} vanishes quickly, and the last two terms dominate $\Sigma_{\text{diff}}$. To reduce $\Sigma_{\text{diff}}$, it suffices to make $\text{Tr}(V\Sigma V^{\top}\Phi)$ small. To this end, we introduce the regularizer 
	\begin{equation}\label{equ:reg}
		\Omega(V):=\text{Tr}(V\Sigma V^{\top}\Phi)
	\end{equation} 
	to the covariance-parameterized LQR problem \eqref{prob:equiV} 
	\begin{equation}\label{prob:regu}
		\begin{aligned}
			&\mathop{\text {minimize}}\limits_{V, \Sigma\succeq 0}~ J_{\lambda}(V):= J(V) + \lambda\Omega(V),\\
			&\text{subject to}~ ~\Sigma = I_n + \overline{X}_1V\Sigma V^{\top}\overline{X}_1^{\top},\overline{X}_0V= I_n
		\end{aligned}
	\end{equation}
	with the gain matrix $K = \overline{U}_0V$ and the regularization coefficient $\lambda>0$. We refer to \eqref{prob:regu} as \textit{regularized covariance parameterization} of the LQR problem. 
	\begin{remark}
		In the prior parameterization \eqref{equ:relation} and \eqref{prob:equi}, the robustness-promoting regularizer $\text{Tr}(G\Sigma G^{\top})$ is adopted to reduce the difference in the Lyapunov equation~\cite{dorfler2021certainty}. By leveraging the relation \eqref{equ:rela} and removing the nullspace, we have $\text{Tr}(G\Sigma G^{\top}) = \text{Tr}(V\Sigma V^{\top}\Phi)/t$. That is, the regularized covariance parameterization problem \eqref{prob:regu} is equivalent to the problem \eqref{prob:equi} with blended (infinite-coefficient) certainty-equivalence promoting regularizer $\|\Pi_{D_0}G \|$  and robustness-promoting regularizer  $\text{Tr}(G\Sigma G^{\top})$. \qed
	\end{remark}
	
	\begin{remark}   \label{remark:regu}
	The regularization coefficient $\lambda$ should be set proportionally to  $1/\sqrt{t}$. To see this, note that $\Sigma_{\text{diff}}$ in \eqref{equ:gap} can be approximated as 
	$$
	\Omega(V)\cdot \frac{\text{constant}}{\sqrt{t}} + o\left(\frac{1}{\sqrt{t}}\right).
	$$
	This means that the uncertainty in $\Sigma$ decreases as $\mathcal{O}(1/\sqrt{t})$, and accordingly, the regularizer reflecting this uncertainty should also diminish at the rate $\mathcal{O}(1/\sqrt{t})$. \qed
	\end{remark}

%
%
	
	Next, we show that the regularized problem \eqref{prob:regu} can be formulated as a convex program. By the change of variables $V = S \Sigma^{-1}$,  \eqref{prob:regu} becomes  
	\begin{align*}
	& \min _{(\Sigma, S, L, M)} \text{Tr}(Q\Sigma) + \text{Tr}(RL) + \lambda \text{Tr}(M\Phi) \\
	& \text { subject to }  
	  \left\{\begin{array}{l}
	  \overline{X}_0 S=\Sigma \\
	  \Sigma \succeq I_n \\
	\overline{X}_1S\Sigma^{-1}S^{\top}\overline{X}_1^{\top} - \Sigma +I_n \preceq 0 \\
	L-\overline{U}_0 S \Sigma^{-1} S^{\top} \overline{U}_0^{\top} \succeq 0 \\
	M- S \Sigma^{-1} S^{\top} \succeq 0.
	\end{array}\right. 
	\end{align*}
	Using a Schur complement, it can be further formulated as an SDP with linear matrix inequality constraints
	\begin{equation}\label{prob:sdp}
	\begin{aligned}
	& \min _{(\Sigma, S, L, M)} \text{Tr}(Q\Sigma) + \text{Tr}(RL) + \lambda \text{Tr}(M\Phi) \\
	& \text { subject to } \\
	& \left\{\begin{array}{l}
	 \overline{X}_0 S=\Sigma, ~~~~~~~~~
	\begin{bmatrix}
	\Sigma - I_n & \overline{X}_1S \\
	S^{\top}\overline{X}_1^{\top} & \Sigma
	\end{bmatrix} \succeq 0 \\
	 \begin{bmatrix}
	 L & \overline{U}_0S \\
	 S^{\top}\overline{U}_0^{\top} & \Sigma
	 \end{bmatrix}\succeq 0, ~~~
	\begin{bmatrix}
	M &S \\
	S^{\top} & \Sigma
	\end{bmatrix} \succeq 0
	\end{array}\right. 
	\end{aligned}
	\end{equation}
	with the control gain $K = \overline{U}_0 S \Sigma^{-1}$. Since the dimension of \eqref{prob:sdp} does not depend on the data length, it can be solved efficiently by modern SDP solvers (e.g., cvx \cite{cvx}).

	\subsection{Harnessing uncertainty in the cost via regularization}
 	We show that the regularizer $\Omega(V)$ can also be used to reduce the uncertainty in the cost function induced by noise. Recall that the covariance-parameterized cost is $$
 	J(V) = \text{Tr}\left((Q+V^{\top}\overline{U}_0^{\top}R\overline{U}_0V)\Sigma\right),
 	$$
 	where the steady-state covariance matrix $\Sigma$ is given by the solution to \eqref{equ:lyap}. Recall also that the true cost that we aim to minimize is
 	$$
 	C(K) = \text{Tr}\left((Q+K^{\top}RK)\Sigma\right),
 	$$
 	where $K = \overline{U}_0V$,
 	 and the steady-state covariance matrix $\Sigma$ is given by the solution to \eqref{equ:true_lyap}. By approximating the difference between those two  steady-state covariance matrices as $\Sigma_{\text{diff}}$, the uncertainty in the cost can be approximated as
 	$$
 	C(K) - J(V) \approx \text{Tr}\left((Q+V^{\top}\overline{U}_0^{\top}R\overline{U}_0V)\cdot \Sigma_{\text{diff}} \right).
 	$$
 	That is, the uncertainty in the cost depends on the multiplication of two terms: the first term $Q+V^{\top}\overline{U}_0^{\top}R\overline{U}_0V$ contains penalty matrices $Q$ and $R$ related to the control goal, and the second term $\Sigma_{\text{diff}}$ \eqref{def:diff} is the uncertainty that affects the closed-loop stability of the solution. Thus, by properly selecting the coefficient  (depending on $Q$ and $R$), the regularizer $\Omega(V)$ in \eqref{equ:reg} accounting for $\Sigma_{\text{diff}}$ can also minimize the uncertainty in the cost. As noted in Remark \ref{remark:regu}, the regularization coefficient here should also be set proportionally to $1/\sqrt{t}$.
 	
 	Our regularization method is also in line with the separation principle literature in data-driven predictive control~\cite{chiuso2023harnessing, grimaldi2024bayesian}, where the covariance of model estimator is added to the cost function to account for the uncertainty. To see this, we interpret our regularization method in terms of indirect control. Leveraging the covariance parameterization in \eqref{equ:newpara}, the regularizer can be reformulated as a function of $K$  
 	\begin{equation}\label{equ:regg}
 	 	\text{Tr}(V\Sigma V^{\top}\Phi) = \text{Tr}\left(\Phi^{-1} \begin{bmatrix}
 		K \\
 		I_n
 	\end{bmatrix} \Sigma \begin{bmatrix}
 		K \\
 		I_n
 	\end{bmatrix}^{\top} \right).
 	\end{equation}
 	Thus, our regularization is also linear in the covariance of the least-squares estimator $\Phi^{-1}/t$ in \eqref{equ:variance} as in \cite{chiuso2023harnessing, grimaldi2024bayesian}.


 	\subsection{Connection to exploration and exploitation in reinforcement learning and indirect adaptive LQR}
 		By using trace properties, the LQR cost $C(K)$ in \eqref{equ:transfer} can be written as
 	\begin{align*}
 		\text{Tr}((Q + K^{\top}RK)\Sigma) = \text{Tr}\left( \begin{bmatrix}
 			R & 0 \\
 			0 & Q
 		\end{bmatrix} \begin{bmatrix}
 			K \\
 			I_n
 		\end{bmatrix} \Sigma \begin{bmatrix}
 			K \\
 			I_n
 		\end{bmatrix}^{\top}  \right).
 	\end{align*}
 	Further, adding the regularizer in the form of \eqref{equ:regg} to the cost function leads to the regularized LQR cost
 	\begin{equation}\label{equ:reg_cost}
 		\text{Tr}\left( \left(\begin{bmatrix}
 			R & 0 \\
 			0 & Q
 		\end{bmatrix} + \lambda\Phi^{-1} \right) \begin{bmatrix}
 			K \\
 			I_n
 		\end{bmatrix} \Sigma \begin{bmatrix}
 			K \\
 			I_n
 		\end{bmatrix}^{\top}  \right).
 	\end{equation}
 	It follows from \eqref{equ:reg_cost} that the regularizer can be interpreted as a correction to the penalty matrices of the LQR problem. 
 	
 	With a positive or negative coefficient $\lambda$, the regularizer can  promote either exploitation or exploration, which are a well-known trade-off in reinforcement learning~\cite{recht2019tour}. Let us elaborate. Since $\Phi^{-1}$ quantifies the covariance of model error, its eigenvector corresponding to large eigenvalue describes the ``direction" of the system with large uncertainty. If $\lambda > 0$, this regularization term in the penalty matrix will punish the system for generating data in the direction with large uncertainty, i.e., it tends to select safe actions that are believed to lead to lowest cost based on the certain part of the system. Thus, the regularizer with positive coefficient promotes exploitation. If $\lambda < 0$, this regularization term will encourage the system for generating data in the direction with large uncertainty, i.e., encourage the system to explore the uncertain part of the system even though it does not minimize the cost immediately. That is, negative coefficient promotes exploration and helps reduce the uncertainty.
 	
 	There are references in indirect adaptive LQR literature that adopt such a regularizer \eqref{equ:reg_cost} for efficient exploration, while the derivation and motivation are different and independent of this paper~\cite{cohen2019learning, abeille2020efficient, chekan2024fully}. Specifically, they regard the regularizer as a constraint and use Lagrangian duality method to formulate the augmented LQR cost \eqref{equ:reg_cost} with a negative $\lambda$. We refer to \cite{cohen2019learning, abeille2020efficient, chekan2024fully} for more details.

	\renewcommand\arraystretch{1.5}
	\begin{table*}[htbp]\label{table}
		\begin{center}
			\begin{tabular}{|c|c|c|c|c|c|}
				\hline
				& $\lambda = 0$ & $\lambda =0.01$ & $\lambda =0.1$ & $\lambda =1$ & $\lambda =10$ \\
				\hline
				\tabincell{c}{$\sigma=0.1$ \\(SNR$\in[5,10]$dB)} & \tabincell{c}{\bm{$\mathcal{S} = 100\%$} \\ \bm{$\mathcal{M} = 0.0042$}} & \tabincell{c}{{$\mathcal{S} = 100\%$ }\\ {$\mathcal{M} = 0.0044$}}  & \tabincell{c}{$\mathcal{S} = 100\%$ \\ $\mathcal{M} = 0.0164$}&  \tabincell{c}{{$\mathcal{S} = 100\%$} \\ $\mathcal{M} = 0.1155$} & \tabincell{c}{$\mathcal{S} = 100\%$ \\ $\mathcal{M} = 0.2168$}\\
				\hline
				\tabincell{c}{$\sigma=0.3$ \\(SNR$\in[0,5]$dB)} & \tabincell{c}{$\mathcal{S} = 100\%$ \\ $\mathcal{M} = 0.039$} & \tabincell{c}{\bm{$\mathcal{S} = 100\%$} \\ \bm{$\mathcal{M} = 0.038$}}  & \tabincell{c}{$\mathcal{S} = 100\%$ \\ $\mathcal{M} = 0.041$}&  \tabincell{c}{{$\mathcal{S} = 100\%$} \\ $\mathcal{M} = 0.142$} & \tabincell{c}{$\mathcal{S} = 100\%$ \\ $\mathcal{M} = 0.243$}\\
				\hline
				\tabincell{c}{$\sigma=0.7$ \\(SNR$\in[-3,0]$dB)} & \tabincell{c}{$\mathcal{S} = 88\%$ \\ $\mathcal{M} = 0.2697$} & \tabincell{c}{$\mathcal{S} = 91\%$ \\ $\mathcal{M} = 0.255$} & \tabincell{c}{\bm{$\mathcal{S} = 99\%$} \\ \bm{$\mathcal{M} = 0.189$}} & \tabincell{c}{$\mathcal{S} = 100\%$ \\ $\mathcal{M} = 0.282$} & \tabincell{c}{$\mathcal{S} = 99\%$ \\ $\mathcal{M} = 0.370$}\\
				\hline
				\tabincell{c}{$\sigma=1$ \\(SNR$\in[-5,-3]$dB)} & \tabincell{c}{$\mathcal{S} = 78\%$ \\ $\mathcal{M} = 0.551$} & \tabincell{c}{$\mathcal{S} = 81\%$ \\ $\mathcal{M} = 0.521$} & \tabincell{c}{{$\mathcal{S} = 97\%$} \\ {$\mathcal{M} = 0.419$}} & \tabincell{c}{\bm{$\mathcal{S} = 97\%$} \\ \bm{$\mathcal{M} = 0.418$}} & \tabincell{c}{$\mathcal{S} = 96\%$ \\ $\mathcal{M} = 0.531$}\\
				\hline
			\end{tabular}
		\end{center}
		\caption{Comparison for different $\lambda$ and noise variance $\sigma^2$ under the offline setting.}
	\end{table*}

\begin{figure}[t] 
	\centerline{\includegraphics[width=80mm]{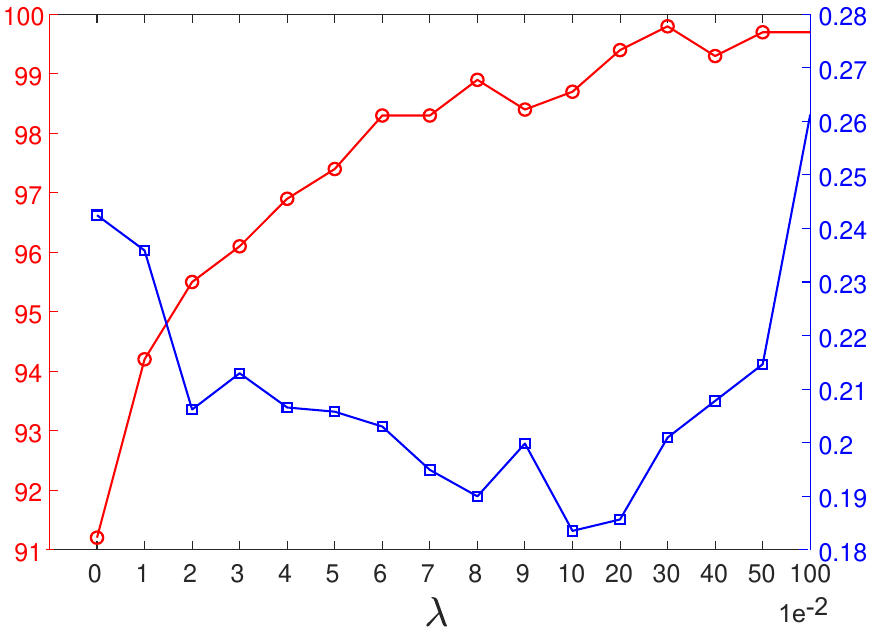}}
	\caption{Performance of the regularized covariance-parameterized LQR \eqref{prob:regu} as a function of $\lambda$. The red line represents percentage of stabilizing solution from $1000$ independent trials, and the blue line represents the median of optimality gap \eqref{equ:error}. The case $\lambda = 0$ corresponds to the certainty-equivalent solution of \eqref{prob:indirect} and \eqref{prob:equiV}.}
	\label{pic:figure}
\end{figure}
	
\section{Simulation}\label{sec:simu}
This section uses simulations to validate our regularization method for the covariance-parameterized LQR. Our simulation is based on the benchmark LQR problem with the system \cite[Section 6]{dean2020sample}
\begin{equation}\label{equ:benchmark}
	\begin{aligned}
		&A = \begin{bmatrix}
			1.01 &  0.01  &  0\\
			0.01  & 1.01 &  0.01\\
			0  &  0.01  &  1.01
		\end{bmatrix}, ~B = I_3,
	\end{aligned}
\end{equation}
which corresponds to a discrete-time marginally unstable Laplacian system. Let $Q= I_3$ and $R = 10^{-3} \times I_3$. This parameter setting is also consistent with \cite{dorfler2021certainty,dorfler22on}.

Consider multiple independent trials. For each trial, we run an experiment on the system with   $x_t \sim \mathcal{N}(0,I_n)$, input $u_t \sim \mathcal{N}(0,I_m)$, and noise $w_t \sim \mathcal{N}(0,\sigma^2\times I_n)$ for time horizon $T = 20$.  We let $K(k)$ be the controller obtained in $k$-th trial by solving the SDP \eqref{prob:sdp}. Whenever $K(k)$ is stabilizing, define the empirical optimality gap
\begin{equation}\label{equ:error}
\mathcal{E}_k = \frac{C(K(k)) - C^*}{C^*}.
\end{equation}
We denote by $\mathcal{S}$ the percentage of finding a stabilizing controller and by $\mathcal{M}$ the median of $\mathcal{E}_k$ through all trials that return a stabilizing controller. The signal-to-noise ratio (SNR) is approximated by $\underline{\sigma}(D_0)/\|W_0\|$ as in \cite{dorfler2021certainty}.

Figure \ref{pic:figure} shows the results from $1000$ independent trails under the noise level $\sigma = 0.7$ (SNR$\in [-3,0]$dB). It can be observed that, compared to the certainty-equivalent solution \eqref{prob:indirect} and \eqref{prob:equiV}, regularization both increases the percentage of obtaining stabilizing solution and reduces the optimality gap for all $\lambda \in (0, 1)$.  

A detailed report for different $\sigma$ from $100$ independent trails is contained in Table I, where we use bold font to highlight the best overall performance in each row. It can be observed that, even for the unreasonable noise level $\sigma = 1$, $97\%$ of the solutions are stabilizing. The optimal selection of the regularization coefficient depends on the SNR, or the noise variance $\sigma^2$ here: a larger SNR corresponds to a smaller uncertainty and hence a small coefficient.
 
\section{Conclusions}\label{sec:conc}
In this paper, we have proposed a regularization method for the covariance parameterization for direct data-driven LQR control. We have shown that the regularizer can well account for the uncertainty both in the steady-state covariance matrix and in the cost functions.  It is worth noting that our approach is simple both in derivation and implementation.

While our simulation validates the effectiveness of the proposed regularization method, a rigorous theoretical understanding of the performance remains open. A comparison study to indirect LQR with regularized system identification method~\cite{pillonetto2022regularized} rather than ordinary least-squares is an interesting future work. It would be also valuable to compare with the robust control methods (e.g., the data informativity approach \cite{van2020data,van2020noisy}) in terms of robust closed-loop stability. Applying the regularization to adaptive control (e.g., data-enabled policy optimization~\cite{zhao2024data}) is an interesting direction.   

\section*{Acknowledgment}
We thank Lenart Treven from ETH Z\"{u}rich for fruitful discussions on the interpretation of the regularizer.
	
\bibliographystyle{IEEEtran}
\bibliography{mybibfile}
\end{document}